\documentclass[pra,twocolumn,showpacs]{revtex4}
\usepackage[dvipdf]{graphicx}
\usepackage{amsmath}
\usepackage{amsfonts}
\usepackage{amssymb}

\providecommand{\U}[1]{\protect\rule{.1in}{.1in}}
\begin{document}

\title{Single Atoms on an Optical Nanofiber}
\author{K. P. Nayak and K. Hakuta}
\affiliation{Department of Applied Physics and Chemistry, University of
Electro-Communications, Chofu, Tokyo 182-8585, Japan}
\date{\today }

\begin{abstract}
We show that single-atoms can be trapped on the surface of a
subwavelength-diameter silica-fiber, an optical nanofiber, without any
external field, and that single photons spontaneously emitted from the atoms
can be readily detected through the single guided-mode of the nanofiber. A
key point of the work is our finding that atom trapping sites are created on
the nanofiber surface by irradiating the atom cloud around the nanofiber
with a violet laser radiation.
\end{abstract}

\pacs{42.50.-p,32.80.Pj,42.62.Fi,32.80.Qk}
\maketitle

In recent years significant progress has been achieved in manipulating
single atoms. Many kinds of single-atom-localizing and trapping schemes have
been demonstrated so far using various external fields. Examples would be
magneto-optical traps with high magnetic field gradient \cite{one,two,three}%
, ion traps \cite{four,five}, far-off-resonance optical traps \cite%
{six,seven} or high-finesse optical cavities \cite{eight,nine}. However, in
order to further extend these technologies, development of a simpler
atom-trapping method without external fields will prove promising. Also in
this context a major problem arises of how to detect the small number of
atoms. Quite often, these experiments rely on complicated design of high
numerical aperture optics or high-finesse optical cavities that may add to
the technical difficulties. Hence development of a more realistic detection
system will also accelerate the technological advancements.

\begin{figure}[ptb]
\begin{center}
\includegraphics[width=8cm]{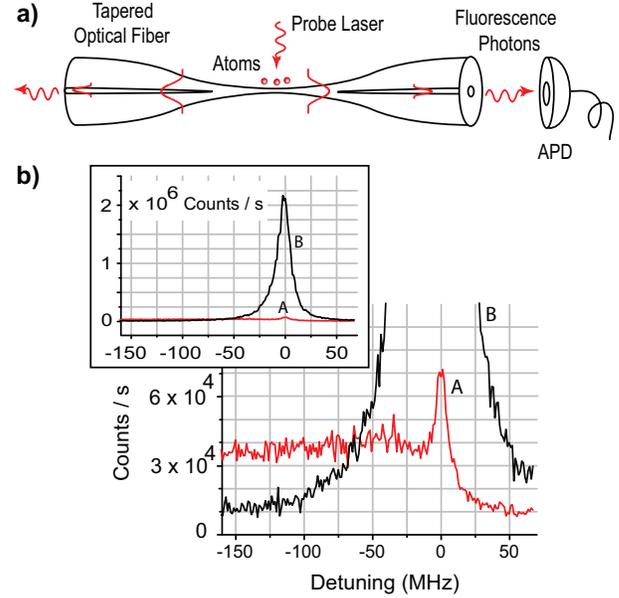}
\end{center}
\caption{(a) Conceptual diagram of the experiment. The nanofiber locates at
the waist of a tapered optical fiber. Fluorescence photons coupled to the
guided mode of the nanofiber are detected at one end of the single mode
optical fiber using an avalanche photodiode (APD). (b) Fluorescence
excitation spectra measured through the nanofiber for the closed cycle
transition, 6S$_{1/2}$ F = 4 $\leftrightarrow$ 6P$_{3/2}$ F' = 5. Detuning
is measured with respect to the atomic resonance. Traces A and B correspond
to without and with the effect of violet laser irradiation, respectively.
The spectra are displayed in two vertical scales so that the change of the
spectrum is readily seen.}
\label{fig1}
\end{figure}

Here we show that single atoms can be trapped on a subwavelength-diameter
silica-fiber, an optical nanofiber \cite{ten,eleven}, and that single
photons spontaneously emitted from the atoms can be readily guided into a
single-mode optical fiber. Experiments are performed by overlapping cold
Cs-atoms in a magneto-optical trap (MOT) with an optical nanofiber and
observing fluorescence photons through the nanofiber after switching off the
MOT laser beams. A key point of the work is our finding that atom trapping
sites are created on the nanofiber surface by irradiating the atom cloud
around the nanofiber with a violet laser radiation.

Figure 1(a) shows the conceptual diagram of the work. The nanofiber is
located at the waist of a tapered optical fiber which is produced by heating
and pulling commercial single-mode optical fibers. The fibers are
adiabatically tapered so that the single mode propagation condition can be
maintained for the whole fiber length. Note that atoms around the nanofiber
emit an appreciable fraction of the fluorescence photons into the guided
mode, since the mode distribution around the nanofiber is strongly confined
to the guided mode \cite{eleven, twelve}. In the present experiments, we use
nanofibers \cite{eleven} with a diameter of 400 nm and a length of 2 mm. A
MOT equipped with a resistively heated alkali dispenser source is used to
produce cold Cs-atoms. The MOT position is controlled to overlap with the
nanofiber \cite{eleven}. Atom number density in MOT and MOT size are
controlled by adjusting the dispenser current. The maximum number density
around the nanofiber is 7 $\times $ 10$^{9}$ cm$^{-3}$ with a MOT diameter
of 2 mm and the temperature of atoms is around 100 $\mu $K. The MOT laser
beams are switched off for 10 $\mu $s periodically at an interval of 200 $%
\mu $s. During the switched-off periods, atoms around the nanofiber are
excited by a probe laser beam irradiated perpendicular to the nanofiber in a
standing-wave configuration with a polarization perpendicular to the fiber.
Fluorescence photons are observed at one end of the fiber using an avalanche
photodiode and a single-photon counting system, and the photon counts are
accumulated for many cycles.

Figure 1(b) shows the excitation spectra measured for a closed cycle
transition, 6S$_{1/2}$ F = 4 $\leftrightarrow $ 6P$_{3/2}$ F' = 5, by
scanning the probe laser frequency around the atomic resonance. The MOT is
set to the maximum density condition and the probe laser diameter is set to
2 mm to irradiate the whole atom cloud. Intensity of the probe laser is set
to 3.3 mW/cm$^{2}$. The trace A denotes the observed spectrum without the
effect of violet laser. As reported previously, the observed line shape is
quite different from the usual atomic line shape \cite{eleven, thirteen,
fourteen}. The spectrum consists of a sharp peak near the resonance and a
long tail on the red detuned side. These observations are attributed to the
van der Waals (vdW) interaction between the Cs-atom and the nanofiber
surface which would be dominant for distances closer than $\lambda /2\pi $
from the surface. The observed spectrum is well understood through a process
in which atoms close to the nanofiber fall into a deep vdW potential. During
the experiments, we have found that the excitation spectrum changes
drastically when we have irradiated the nanofiber with violet laser
radiation of wavelength 407 nm. The conditions are the following: The
nanofiber region is irradiated with the violet laser in the presence of the
MOT for several minutes. Irradiating power is 5 mW with a beam diameter of 2
mm (the irradiating intensity is 150 mW/cm$^{2}$). After switching off the
violet laser the fluorescence excitation spectrum is measured. The observed
spectrum is denoted by trace B in Fig. 1(b) and the inset. As seen readily,
the spectrum is very different from the trace A. The spectrum becomes very
much squeezed towards the atomic resonance, and the peak fluorescence count
increases almost 30 times. The width of the spectrum is 15 MHz FWHM. We
should note that the irradiation in the absence of the MOT has no effect on
the spectrum. We have found that once the nanofiber is irradiated, the
observed effect lasts for several days. The observations suggest that the
violet laser irradiation has modified the nanofiber surface in such a way
that most of the atoms are kept from falling into the vdW potential, and
dwell near the surface at some specific distance.

In order to clarify the single-atom characteristics around the nanofiber, we
perform Hanbury-Brown and Twiss (HBT) experiments by reducing the atom
number. The fluorescence light through fiber are split into two using a 3 dB
fiber coupler, and are detected by two separate avalanche photodiodes. The
photon correlations between the two channels are measured using a
time-correlated single-photon-counter with a time resolution of 1 ns. In the
measurements, the dispenser current is decreased to reduce both atom number
and MOT size. The minimum MOT size is 80 $\mu $m in diameter with an atom
density of $\sim $\textit{\ }$0.7\times 10^{9}$ cm$^{-3}$ for dispenser
current I$_{D}$ = 3.8 A. The probe laser is tuned close to the atomic
resonance, and is line focused down to 100 $\mu $m using a cylindrical lens
to spatially restrict the observation region to the atom cloud. The focused
probe laser intensity is 60 mW/cm$^{2}$. Average atom number in the
observation region is estimated to be much less than one for I$_{D}$ = 3.8
A, assuming an observation volume around the nanofiber of 200 nm in
thickness \cite{eleven} and 100 $\mu $m in length. Fluorescence photon count
is  $\sim $\textit{\ }$1\times 10^{4}$ counts/s under this condition.

\begin{figure}[tbp]
\begin{center}
\includegraphics[width=8cm]{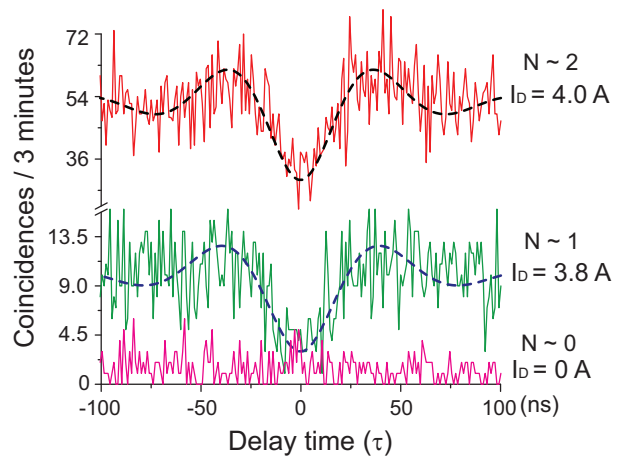}
\end{center}
\caption{Photon correlations with Hanbury-Brown and Twiss arrangement for
atoms around the nanofiber. $\protect\tau $ and I$_{D}$ denote the delay
time between the two channels and the dispenser current, respectively.
Dashed curves denote the theoretically calculated photon coincidences for N
= 1 and 2, respectively, where N denotes the number of atoms.}
\label{fig2}
\end{figure}

Figure 2 displays the observed coincidences for different delay time between
the two channels. Each curve is obtained after an integration time of 3
minutes which requires a measurement time of 1 hour. The coincidences for I$%
_{D}$ = 0 A correspond to the background for the uncorrelated scattered
light from the nanofiber. The coincidences for I$_{D}$ = 3.8 A and 4 A show
the correlations for the fluorescence photons. Both curves show clear
antibunching of fluorescence photons at zero time-delay and Rabi oscillation
behaviors in the wings. The curves are fitted using a formula for photon
coincidences from N-atoms \cite{eighteen}, $Ng^{(2)}(\tau )+N(N-1)$, where $%
g^{(2)}(\tau )$ is the correlation function for a single atom and is
calculated assuming spontaneous emission time of 30 ns and Rabi frequency of
13 MHz. Observed and fitted curves are in good agreement. Thus, observed
coincidences for I$_{D}$ = 3.8 A and 4 A are ascribed to photon correlations
from one atom and two atoms, respectively. Note that the antibunching is
observable for atoms only after the violet laser irradiation. We should note
that the assumed Rabi frequency is 1.3 times smaller in value than that
simply estimated for the F = 4 $\leftrightarrow $ 5 transition.

\begin{figure}[ptb]
\begin{center}
\includegraphics[width=7cm]{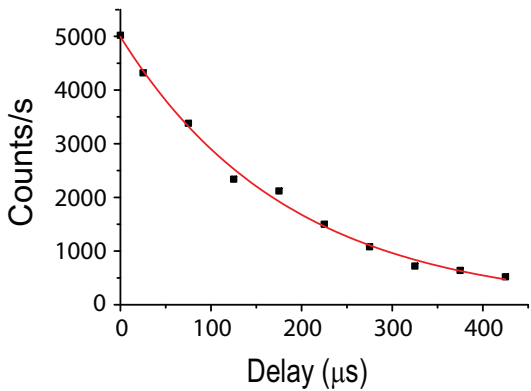}
\end{center}
\caption{Temporal behavior of fluorescence signal measured through the
nanofiber under single-atom condition. The black dots denote the observed
fluorescence counts at various delay time after switching off the MOT beams.
The red curve is the exponential fit to the data giving a decay time of 180 $%
\protect\mu $s.}
\label{fig3}
\end{figure}

Next, the temporal behavior of the fluorescence signals is measured under
the single-atom condition. Measurements are performed by extending the
switched-off (on) periods to 1 ms (20 ms). Signals are measured for a gate
time of 50 $\mu $s in the switched-off periods, at various delay time after
switching off the MOT beams. The signals are accumulated for many cycles.
The results are exhibited in Fig. 3. The temporal behavior shows an
exponential decay with a lifetime of 180 $\mu $s. This lifetime corresponds
to the atom dwelling time in the observation volume. If atoms are free
around the nanofiber, the dwelling time may be determined by the transit
time across the observation volume. Assuming a mean velocity of atoms as 10
cm/s and an atom transit-length of 1 $\mu $m, one can estimate the transit
time to be $\sim $ 10 $\mu $s, which is about twenty-times shorter than the
observed lifetime. It means that atoms around the nanofiber are not
completely free; atoms are localized in the close vicinity of the nanofiber
surface.

\begin{figure}[ptb]
\begin{center}
\includegraphics[width=7cm]{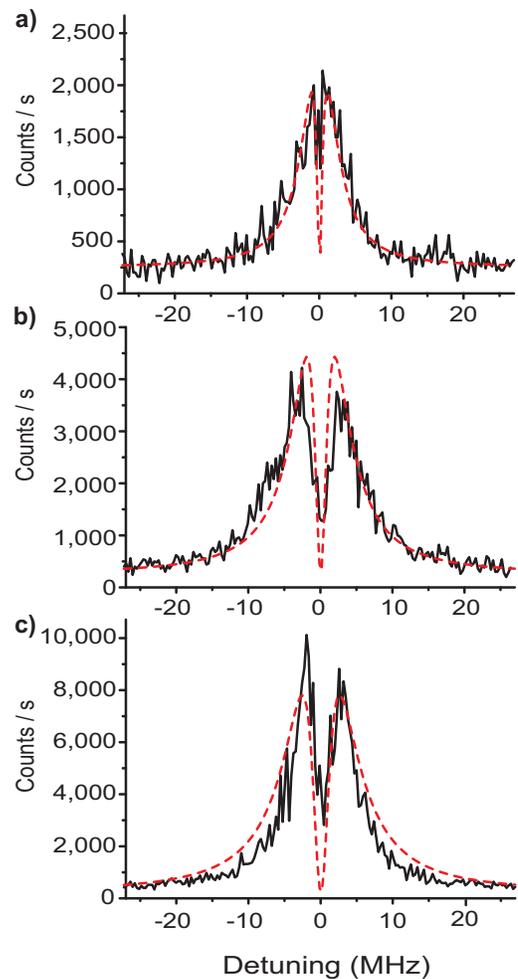}
\end{center}
\caption{Solid curves in (a), (b), and (c) exhibit the fluorescence
excitation spectra measured through the nanofiber under one single-atom
condition for three different probe laser intensities 0.7, 3.5 and 7 mW/cm$%
^{2}$, respectively. Detuning is measured with respect to the atomic
resonance. Dashed curves are theoretically calculated spectra assuming a
V-type three level scheme.}
\label{fig4}
\end{figure}

We measure the excitation spectra under the single atom condition to further
clarify the behaviors of atoms. Observed results for three different probe
laser intensities are displayed in Figs. 4(a-c) by solid curves.
Measurements are performed by using the similar procedures as those for Fig.
1(b). Peculiar features are readily seen. For the lowest laser intensity,
the spectrum exhibits almost a Lorentzian shape with 8 MHz FWHM, slightly
broader than the spontaneous-emission lifetime broadening, but a sharp,
small dip is seen at the peak. With increasing laser intensity, the signal
becomes stronger and the sharp dip becomes more apparent, but the dip width
is narrower than the lifetime broadening. Such spectral characteristics with
a sharp dip cannot be explained by simply overlapping two Lorentzian
profiles. Such characteristics can be explained by incorporating a quantum
interference effect. The observed spectral characteristics may be explained
with a V-type three-level scheme discussed in Refs. \cite{ninteen, twenty}.
The scheme consists of two upper levels closely spaced within the radiative
broadening and one lower ground level, and the upper levels spontaneously
decay to the lower level with the same rate. By solving density matrix
equations under stationary condition, the excitation spectrum can be
calculated for various probe laser intensities (Rabi frequencies).
Calculated spectra for three Rabi frequencies are overlaid on the observed
ones in Figs. 4(a-c) with dashed lines. Spacing between the upper levels is
assumed to be 1.5 MHz. The sharp dips are reproduced, and the relative
spectral intensities are also well reproduced. Regarding the Rabi
frequencies, we have used 1.4 times smaller values than those for the F = 4 $%
\leftrightarrow $ 5 two-level transition to keep the total decay rate of the
upper state equal to the value for the F = 5 level.

Observed spectra clearly reveal that the upper atomic state of the
transition is split into two levels. The observed splitting may be
understood as a consequence of the localization of atoms into a tiny
potential in the close vicinity of the nanofiber surface; that is, the atom
is trapped in the potential and, moreover, the atomic motion is quantized
resulting in the two motional sub-levels for the excited electronic state.
We suspect that two sub-levels are also created for the ground electronic
state, because the depths for the center dip can be better reproduced by
introducing two sub-levels for the ground state.

The spectral measurements clarify the discrepancy found in the
HBT-experiments that the fitted Rabi frequencies are much smaller than the
two level values, because the localized atom is not a simple two-level atom,
and the two-level estimation is not appropriate anymore. Regarding the
spectrum B in Fig. 1(b), we have not observed any sharp spectral dip. It may
be understood to be due to some inhomogeneous broadening for many localized
atoms in many localizing sites which might have washed out the quantum
interference effects. Number of localized atoms in the spectrum B may be
estimated to be $\sim $\ 200 by comparing the integrated intensities for the
spectrum B and that for one single-atom.

\begin{figure}[tbp]
\begin{center}
\includegraphics[width=6cm]{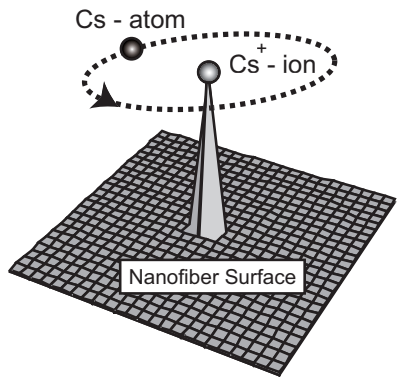}
\end{center}
\caption{Conceptual diagram of atom trapping on a nanofiber surface. A Cs$%
^{+}$-ion stuck at the top of a prominence on the nanofiber surface produces
a Coulomb field so that a dipole moment is induced on a Cs-atom passing near
the ion. Consequently, the Cs-atom orbits around the ion due to the mutual
balance between the Coulomb force on the induced-dipole, the centrifugal
force and the vdW force.}
\end{figure}

We have not yet identified the mechanism of the atom trapping definitively,
but we speculate that the following scenario may be occurring. Regarding the
effect of the violet laser irradiation, it is not due to the light induced
atom desorption (LIAD) well known for alkalis \cite{sixteen}. If it were due
to the LIAD process, the effect would also appear in the absence of the MOT.
By irradiating the MOT atoms with a 407 nm laser, the atoms are
photo-ionized \cite{seventeen} from the excited electronic state 6P$_{3/2}$.
Generated Cs-ions then stick to the cusps of some prominences on the
nanofiber surface, and the stuck charges form atom-trapping orbits around
them due to the mutual balance between the induced-dipole force, the
centrifugal force, and the vdW force, as schematically illustrated in Fig.
5. Since the observed resonances are close to the free atomic resonance and
the width of the trace B in Fig. 1 (b) is 15 MHz, height of the prominences
may distribute from 50 to 100 nm assuming the vdW shift. Similar atom
trapping with orbiting trajectory is discussed in Ref. \cite{orbiting}.
Based on their results and assuming an inverse proportionality of the
orbiting frequency to the orbit radius, one can expect an orbit radius to be
around 30 nm for the observed splitting of 1.5 MHz.

In conclusion, we have found that single-atom localizing sites are created
on a nanofiber surface by irradiating the MOT atoms around the fiber with a
violet laser radiation so that single atoms are trapped without any external
field, and that single photons spontaneously emitted from the atoms are
readily guided into a single-mode optical fiber. The present finding may be
extended to various surfaces and various atomic and molecular species.

We are thankful to Fam Le Kien, Shinichi Watanabe, Makoto Morinaga, Mark
Sadgrove, and Manoj Das for useful discussions and technical assistance.
This work was carried out under the 21st Century COE program on
\textquotedblleft Innovation in Coherent Optical Science\textquotedblright .

\end{document}